\documentclass[11pt,a4paper]{article}
\usepackage{amssymb}
\usepackage{amsfonts}
\usepackage{helvet}
\usepackage{courier}
\usepackage{bm}
\usepackage{amsmath}

\def\ihalf{\textstyle{\frac{i}{2}}}
\def\ifourth{\textstyle{\frac{i}{4}}}
\def\onehalf{\textstyle{\frac{1}{2}}}

\def\onefourth{\textstyle{\frac{1}{4}}}
\def\D{{\mathcal D}{}}
\def\Gammabol{{\stackrel{\circ}{\Gamma}}{}}{}

\def\Abol{{\stackrel{~\circ}{A}}{}}

\def\Rbol{{\stackrel{\circ}{R}}{}}
\def\Lbol{{\stackrel{\circ}{\mathcal L}}{}}
\def\Tbol{{\stackrel{\circ}{T}}{}}

\def\Gammaw{{\stackrel{\bullet~}{\Gamma}}{}}

\def\Rw{{\stackrel{\bullet}{R}}{}}
\def\Qw{{\stackrel{\bullet}{Q}}{}}
\def\jw{{\stackrel{~\bullet}{J}}{}}

\def\Lw{{\stackrel{~\bullet}{\mathcal L}}{}}
\def\Tw{{\stackrel{\bullet}{T}}{}}

\def\Kw{{\stackrel{\bullet}{K}}{}}
\def\Aw{{\stackrel{\bullet}{A}}{}}

\def\Dw{{\stackrel{~\bullet}{\mathcal D}}{}}

\def\sw{{\stackrel{\bullet}{S}}{}}

\def\be{\begin{equation}}
\def\ee{\end{equation}}
\def\ba{\begin{eqnarray}}
\def\ea{\end{eqnarray}}

\begin{document}
\renewcommand{\thefootnote}{\fnsymbol{footnote}}
\begin{center}
{\Large \bf Lorentz Connections and Gravitation}
\vskip 0.5cm
{\bf J. G. Pereira}
\vskip 0.1cm
{\it Instituto de F\'{\i}sica Te\'orica, UNESP-Univ Estadual Paulista \\
Caixa Postal 70532-2, 01156-970 S\~ao Paulo, Brazil}

\vskip 0.8cm
\begin{quote}
{\bf Abstract.~}{\footnotesize The different roles played by Lorentz connections in general relativity and in teleparallel gravity are reviewed. Some of the consequences of this difference are discussed.

}
\end{quote}
\end{center}
\vskip 0.8cm 

\section{Introduction}

A key point of gravitation is that the metric tensor defines neither curvature nor torsion by itself \cite{koba}. As a matter of fact, curvature and torsion require a {\em connection} to be defined, and many different connections, with different curvature and torsion, can be defined on the very same metric spacetime \cite{livro}. How can we determine the relevant connection for gravitation? This is a fundamental question, which has more than one answer. For example, when constructing general relativity, Einstein chose the {\em zero--torsion} Levi--Civita, or Christoffel connection, which is a connection completely specified by the ten components of the metric tensor. In this theory, therefore, torsion is chosen to vanish from the very beginning, and the gravitational field is represented by curvature.

A second possibility would be to choose a {\em zero--curvature} Lorentz connection not related to gravitation, but to inertial effects only. The gravitational theory that emerges from this choice is teleparallel gravity, a gauge theory for the translation group, in which curvature is assumed to vanish from the very beginning. In this theory, the gravitational field turns out to be represented by a translational gauge potential, which appears as the non--trivial part of the tetrad field and gives rise to a non--vanishing torsion, the field strength of the theory. One may wonder why a gauge theory for the translation group, and not for any other group related to spacetime. The answer is related to the source of gravitation: energy and momentum. From Noether's theorem, an instrumental piece of gauge theories \cite{kopo9}, we know that the energy--momentum tensor is conserved provided the source lagrangian is invariant under spacetime translations. If gravity is to be described by a gauge theory with energy--momentum as source, therefore, it must be a gauge theory for the translation group \cite{livro2}.

Although equivalent to general relativity, teleparallel gravity provides a new insight into gravitation. The purpose of these lectures is to explore some of these insights, as well as discuss how this approach could help to answer some old questions permeating general relativity, like for example the energy localizability of the gravitational field and the problem of quantum gravity. 

\section{Linear Frames and Tetrads}
\label{sec:framestetrads}

Spacetime is the common arena on which the four presently known fundamental interactions manifest themselves. Electromagnetic, weak and strong interactions are described by gauge theories involving transformations taking place in {\em internal} spaces, by themselves unrelated to spacetime. The basic setting of gauge theories are the principal bundles, in which a copy of the corresponding gauge group is attached at each point of spacetime --- the base space of the bundle. Gravitation, on the other hand, is deeply linked to the very structure of spacetime. The geometrical setting of gravitation is the tangent bundle, a natural construction always present in any differentiable manifold: at each point of spacetime there is a tangent space attached to it --- the fiber of the bundle, which is seen as a vector space. In what follows we are going to use the Greek alphabet $(\mu, \nu, \rho, \dots = 0,1,2,3)$ to denote indices related to spacetime, and the first letters of the Latin alphabet $(a,b,c, \dots = 0,1,2,3)$ to denote indices related to the tangent space, a Minkowski spacetime whose Lorentz metric, in cartesian coordinates, is assumed to have the form
\be
\eta_{ab} = \mathrm{diag}(+1,-1,-1,-1).
\label{eq:etaofMinko}
\ee

A general spacetime is a 4-dimensional differential manifold, indicated ${\mathbb R}^{3,1}$, whose tangent space is, at each point, a Minkowski spacetime. Spacetime coordinates will be denoted by $\{x^\mu\}$, whereas tangent space coordinates will be denoted by $\{x^a\}$. Such coordinate systems determine, on their domains of definition, local bases for vector fields, formed by the sets of gradients
\be
\{\partial_\mu\} \equiv \{ {\partial}/{\partial x^\mu} \} \quad \mbox{and} \quad
\{\partial_a\} \equiv \{ {\partial}/{\partial x^a} \},
\ee
as well as bases $\{dx^\mu\}$ and $\{dx^a\}$ for covector fields, or differentials. These bases are dual, in the sense that
\be
dx^\mu \, ({\partial_\nu}) = \delta^\mu_\nu \quad \mbox{and} \quad
dx^a \, ({\partial_b}) = \delta^a_b.
\ee
On the respective domains of definition, any vector or covector can be expressed in terms of these bases, which can furthermore be extended by direct product to constitute bases for general tensor fields of any order.

\subsection{Trivial Frames}
\label{sec:frames}

Trivial frames, or trivial tetrads,\index{Frame!field} will be denoted by 
\be
\{e_{a}\} \quad \mbox{and} \quad \{e^{a}\}.
\ee
The above mentioned coordinate bases 
\be
\{{e_a}\} = \{{\partial_a}\} \quad \mbox{and} \quad
\{{e^a}\} = \{{d x^a}\}
\ee
are very particular cases, whose name stems from their relationship to a coordinate system. Any other set of four linearly independent fields $\{e_{a}\}$ will form another basis, and will have a dual $\{e^{a}\}$ whose members are such that
\be
e^{a}(e_b) = \delta^a_b.
\label{OrtoLiFra}
\ee
Notice that, on a general manifold, vector fields are (like coordinate systems) only locally defined~---~and linear frames, as sets of four such fields, are only defined on restricted domains.

These frame fields are the general linear bases on the spacetime differentiable manifold  ${\mathbb R}^{\,3,1}$. The whole set of such bases, under conditions making of it also a differentiable manifold, constitutes the {\em bundle of linear frames}. A frame field provides, at each point $p \in {\mathbb R}^{\,3,1}$, a basis for the vectors on the tangent space ${T}_p{\mathbb R}^{\,3,1}$. Of course, on the common domains they are defined, each member of a given base can be written in terms of the members of any other. For example,
\be
e_a = e_a{}^\mu \, \partial_\mu \quad \mbox{and} \quad e^{a} = e^{a}{}_\mu \, dx^\mu,
\ee
and conversely,
\be
\partial_\mu = e^a{}_\mu \, e_a \quad \mbox{and} \quad dx^\mu = e_{a}{}^\mu \, e^{a}.
\label{eq:partialmu}
\ee
On account of the orthogonality conditions (\ref{OrtoLiFra}), the frame components satisfy 
\begin{equation}
e^{a}{}_{\mu} e_{a}{}^{\nu} = \delta_{\mu}^{\nu} \quad \mbox{and} \quad
e^{a}{}_{\mu} e_{b}{}^{\mu} = \delta^{a}_{b}.
\label{eq:frameprops1}
\end{equation}
Notice that these frames, with their bundles, are constitutive parts of spacetime: they are automatically present as soon as spacetime is taken to be a differentiable manifold.

A general linear base $\{e_{a}\}$ satisfies the commutation relation
\begin{equation}
[e_{a}, e_{b}] = f^{c}{}_{a b} \; e_{c},
\label{eq:comtable0}
\end{equation}
with $f^{c}{}_{a b}$ the so--called structure coefficients, or coefficients of anholonomy, or still the anholonomy of frame $\{e_{a}\}$. As a simple computation shows, they are defined by
\begin{equation}
f^c{}_{a b} = e_a{}^{\mu} e_b{}^{\nu} (\partial_\nu
e^c{}_{\mu} - \partial_\mu e^c{}_{\nu} ).
\label{fcab0}
\end{equation}
A preferred class is that of inertial frames, denoted $e'_a$, those for which
\be
f'^{a}{}_{cd} = 0.
\label{fcabinertial}
\ee
Base $\{e'^{a}\}$ is then said to {\em holonomic}. Of course, all coordinate bases are holonomic. This is not a local property, in the sense that it is valid everywhere for frames belonging to this inertial class.

Consider now the Minkowski spacetime metric, which in cartesian coordinates has the form
\be
\eta_{\mu \nu} = \mathrm{diag}(+1,-1,-1,-1).
\label{eq:etaofMinkoST}
\ee
In any other coordinates, $\eta_{\mu \nu}$ will be a function of the spacetime coordinates. The linear frame
\be
e_{a} = e_{a}{}^{\mu} \, {\partial_{\mu}},
\ee
provides a relation between the tangent--space metric $\eta_{a b}$ and the spacetime metric $\eta_{\mu \nu}$. Such relation is given by
\begin{equation}
\eta_{a b} = {\eta}_{\mu \nu} \, e_{a}{}^{\mu} e_{b}{}^{\nu}.
\label{gtoeta}
\end{equation}
Using the orthogonality conditions (\ref{eq:frameprops1}), the inverse relation is found to be
\begin{equation}
{\eta}_{\mu \nu} = \eta_{a b} \, e^{a}{}_{\mu} e^{b}{}_{\nu}.
\label{eq:tettomet0}
\end{equation}
Independently of whether $e_{a}$ is holonomic or not, or equivalently, whether they are inertial or not, they always relate the tangent Minkowski space to a Minkowski spacetime. These are the frames appearing in special relativity, and are usually called trivial frames --- or trivial tetrads.

\subsection{Nontrivial Frames}
\label{sec:tetrads}

Nontrivial frames, or tetrads, will be denoted by
\be
\{h_{a}\} \quad \mbox{and} \quad \{h^{a}\}.
\ee
They are defined as linear frames whose coefficient of anholonomy is related to both inertial effects {\it and} gravitation. Let us consider a general pseudo--riemannian spacetime with metric components $g_{\mu \nu}$ in the some dual holonomic basis $\{d x^{\mu}\}$. The tetrad field
\be
h_{a} = h_{a}{}^{\mu} \, {\partial_{\mu}} \quad \mbox{and} \quad h^a = h^a{}_\mu dx^\mu,
\ee
is a linear basis that relates $g_{\mu \nu}$ to the tangent--space metric $\eta_{a b}$ through the relation
\begin{equation}
\eta_{a b} = g_{\mu \nu} \, h_{a}{}^{\mu} h_{b}{}^{\nu}.
\label{eq:gtoeta}
\end{equation}
The components of the dual basis members
$h^{a} = h^{a}{}_{\nu} dx^{\nu}$ satisfy
\begin{equation}
h^{a}{}_{\mu} \, h_{a}{}^{\nu} = \delta_{\mu}^{\nu} \quad {\rm and} \quad
h^{a}{}_{\mu} \, h_{b}{}^{\mu} = \delta^{a}_{b},
\label{eq:tetradprops1}
\end{equation}
so that Eq.~(\ref{eq:gtoeta}) has the inverse
\begin{equation}
g_{\mu \nu} = \eta_{a b} \, h^{a}{}_{\mu} h^{b}{}_{\nu}.
\label{eq:tettomet}
\end{equation}
We see from these relations that
\be
h = \det (h^a{}_\mu) = \sqrt{-g} \, ,
\ee
with $g = \det(g_{\mu \nu})$. 

A tetrad basis $\{h_{a}\}$ satisfies the commutation relation
\begin{equation}
[h_{a}, h_{b}] = f^{c}{}_{a b}\, h_{c},
\label{eq:comtable}
\end{equation}
with $f^{c}{}_{a b}$ the structure coefficients, or coefficients of anholonomy, of frame $\{h_{a}\}$. The basic difference in relation to the linear bases $\{e_a\}$ is that now the $f^{c}{}_{a b}$ represent both inertia and gravitation.
As before, the structure coefficients are given by
\begin{equation}
f^c{}_{a b} = h_a{}^{\mu} h_b{}^{\nu} (\partial_\nu
h^c{}_{\mu} - 
\partial_\mu h^c{}_{\nu} ).
\label{fcab}
\end{equation}
Although nontrivial tetrads are, by definition, anholonomic due to the presence of gravitation, it is still possible that {\em locally}, $f^{c}{}_{a b}$ = $0$. In this case, $d h^{a} = 0$, which means that $h^{a}$ is locally a closed differential form. In fact, if this holds at a point $p$, then there is a neighborhood around $p$ on which functions (coordinates) $x^a$ exist such that
\[
h^{a} = dx^a.
\]
We say that a closed differential form is always locally integrable, or exact. This is the case of locally inertial frames, which are always holonomic. In these frames, inertial effects locally compensate gravitation.

\section{Lorentz Connections}
\label{sec:connections}

A {\em Lorentz connection} $A_\mu$, frequently referred to also as {\it spin connection}, is a 1-form assuming values in the Lie algebra of the Lorentz group,
\be
A_\mu = \onehalf \, A^{ab}{}_\mu \, S_{ab},
\ee
with $S_{ab}$ a given representation of the Lorentz generators. As these generators are antisymmetric in the algebraic indices, $A^{ab}{}_\mu$ must be equally antisymmetric in order to be lorentzian. This connection defines the Fock--Ivanenko covariant derivative \cite{fi1,fi2}
\be
\D_\mu = \partial_\mu -  \ihalf \, A^{ab}{}_\mu \, S_{ab},
\label{eq:FockIvanenko} 
\ee
whose second part acts only on the algebraic, or tangent space indices. For a scalar field $\phi$, for example, the generators are
\be
S_{ab} = 0.
\ee
For a Dirac spinor $\psi$, they are given by \cite{dirac1}
\be
S_{ab} = \ifourth \left[\gamma_a, \gamma_b \right],
\label{eq:spinorep}
\ee
with $\gamma_a$ the Dirac matrices. A Lorentz vector field $\phi^c$, on the other hand, is acted upon by the vector representation of the Lorentz generators, matrices $S_{ab}$ with entries \cite{ramond1} 
\be
(S_{ab})^c{}_d = i \left(\eta_{bd} \, \delta_a^c  - \eta_{ad} \, \delta_b^c \right).\label{eq:vecrep}
\ee
The Fock--Ivanenko derivative is, in this case,
\be
\D_\mu \phi^c = \partial_\mu \phi^c + A^{c}{}_{d \mu} \, \phi^d,
\label{VectorFI}
\ee
and so on for any other fundamental field.

On account of the soldered character of the tangent bundle, a tetrad field relates tangent space (or internal) tensors with spacetime (or external) tensors. For example, if $\phi^a$ is an internal, or Lorentz vector, then 
\be
\phi^\rho = h_a{}^\rho \, \phi^a
\label{ixe}
\ee
will be a spacetime vector. Conversely, we can write
\be
\phi^a = h^a{}_\rho \, \phi^\rho.
\label{exi}
\ee
On the other hand, due to its non--tensorial character, a connection will acquire a vacuum, or non--homogeneous term, under the same operation. For example, to each spin connection $A^{a}{}_{b \mu}$, there is a corresponding general linear connection $\Gamma{}^{\rho}{}_{\nu \mu}$, given by
\be
\Gamma^{\rho}{}_{\nu \mu} = h_{a}{}^{\rho} \partial_{\mu} h^{a}{}_{\nu} +
h_{a}{}^{\rho} A^{a}{}_{b \mu} h^{b}{}_{\nu} \equiv
h_{a}{}^{\rho} \, \D_{\mu} h^{a}{}_{\nu},
\label{geco}
\ee
where $\D_{\mu}$ is the covariant derivative (\ref{VectorFI}), in which the generators act on internal (or tangent space) indices only. The inverse relation is, consequently,
\be
A^{a}{}_{b \mu} =
h^{a}{}_{\nu} \partial_{\mu}  h_{b}{}^{\nu} +
h^{a}{}_{\nu} \Gamma^{\nu}{}_{\rho \mu} h_{b}{}^{\rho} \equiv
h^{a}{}_{\nu} \nabla_{\mu}  h_{b}{}^{\nu},
\label{gsc}
\ee
where $\nabla_{\mu}$ is the standard covariant derivative in the connection
$\Gamma^{\nu}{}_{\rho \mu}$, which acts on external indices only. For a spacetime vector $\phi^\nu$, for example, it is given by
\be
\nabla_\mu \phi^\nu = \partial_\mu \phi^\nu + \Gamma^\nu{}_{\rho \mu} \, \phi^\rho.
\ee
Using relations (\ref{ixe}) and (\ref{exi}), it is easy to verify that \cite{kibble}
\be
\D_\mu \phi^d = h^d{}_\rho \, \nabla_\mu \phi^\rho.
\label{iDxeD}
\ee
Equations (\ref{geco}) and (\ref{gsc}) are simply different ways of expressing the property
that the total covariant derivative of the tetrad~---~that is, a covariant derivative with connection terms for both internal and external indices~---~vanishes identically:
\be
\partial_{\mu} h^{a}{}_{\nu} - \Gamma^{\rho}{}_{\nu \mu} h^{a}{}_{\rho} +
A^{a}{}_{b \mu} h^{b}{}_{\nu} = 0.
\label{todete}
\ee

\subsection{Curvature and Torsion}
\label{sec:CurvTor}

Curvature and torsion require a Lorentz connection to be defined \cite{koba}. Given a Lorentz connection $A^{a}{}_{b \mu}$, the corresponding curvature is a 2-form assuming values in the Lie algebra of the Lorentz group,
\be
R_{\nu \mu} = \onehalf \; R^{ab}{}_{\nu \mu} \; S_{ab}.
\ee
Torsion is also a 2-form, but assuming values in the Lie algebra of the translation group, 
\be
T_{\nu \mu} = T^{a}{}_{\nu \mu} \, P_a,
\ee
with $P_a = \partial_a$ the translation generators. The curvature and torsion components are given, respectively, by
\be
R^{a}{}_{b \nu \mu} = \partial_{\nu} A^{a}{}_{b \mu} -
\partial_{\mu} A^{a}{}_{b \nu} + A^a{}_{e \nu} A^e{}_{b \mu}
- A^a{}_{e \mu} A^e{}_{b \nu}
\label{curvaDef}
\ee
and
\be
T^a{}_{\nu \mu} = \partial_{\nu} h^{a}{}_{\mu} -
\partial_{\mu} h^{a}{}_{\nu} + A^a{}_{e \nu} h^e{}_{\mu}
- A^a{}_{e \mu} h^e{}_{\nu}.
\label{tordef}
\ee
Through contraction with tetrads, these tensors can be written in spacetime--indexed forms:
\be
R^\rho{}_{\lambda\nu\mu} = h_a{}^\rho \, h^b{}_\lambda \, R^a{}_{b \nu \mu},
\label{RatoRmi}
\ee
and
\be
T^\rho{}_{\nu \mu} = h_a{}^\rho \, T^a{}_{\nu \mu}.
\label{TatoTmi}
\ee
Using relation (\ref{gsc}), their components are found to be 
\be
\label{sixbm}
R^\rho{}_{\lambda\nu\mu} = \partial_\nu \Gamma^\rho{}_{\lambda \mu} -
\partial_\mu \Gamma^\rho{}_{\lambda \nu} +
\Gamma^\rho{}_{\eta \nu} \Gamma^\eta{}_{\lambda \mu} -
\Gamma^\rho{}_{\eta \mu} \Gamma^\eta{}_{\lambda \nu}
\ee
and
\be
T^\rho{}_{\nu \mu} =
\Gamma^\rho{}_{\mu\nu} - \Gamma^\rho{}_{\nu\mu}.
\label{sixam}
\ee

Since the spin connection $A^a{}_{b \nu}$ is a four--vector in the last index, it satisfies
\be
A^a{}_{bc} = A^a{}_{b \nu} \, h_c{}^\nu.
\ee
It can thus be verified that, in the anholonomic basis $\{h_a\}$, the curvature and
torsion components are given respectively by 
\be
R{}^{a}{}_{b cd } =    
h_c \left(A^a{}_{b d} \right) - h_d \left(A^a{}_{b c} \right) + A^a{}_{e c}A^e{}_{b d} - A^a{}_{e d} A^e{}_{b c} - f^{e}{}_{c d} A^a{}_{b e}  
\label{13bm}
\ee
and 
\be\
T^a{}_{bc} = A^a{}_{cb} - A^a{}_{bc} - f^a{}_{bc}, 
\label{13am}
\ee
where, we recall, $h_c = h_c{}^\mu \partial_\mu$. Use of~(\ref{13am}) for three different combinations of indices gives
\begin{equation}%
A^{a}{}_{b c} = \onehalf (f_{b}{}^{a}{}_{c} + T_{b}{}^{a}{}_{c} 
+ f_{c}{}^{a}{}_{b} + T_{c}{}^{a}{}_{b}
- f^{a}{}_{b c} - T^{a}{}_{b c}).
\label{tobetaken2}
\end{equation}%
This expression can be rewritten in the form
\be
A^a{}_{bc} = \Abol^a{}_{bc} + K^a{}_{bc},
\label{rela0alge}
\ee
where
\begin{equation}%
\Abol^{a}{}_{b c} = \onehalf \left(f_{b}{}^{a}{}_{c} 
+ f_{c}{}^{a}{}_{b} - f^{a}{}_{b c} \right) 
\label{tobetaken30}
\end{equation}%
is the usual expression of the general relativity spin connection in terms of the coefficients of anholonomy, and
\begin{equation}%
K^{a}{}_{b c} = \onehalf \left(T_{b}{}^{a}{}_{c} 
+ T_{c}{}^{a}{}_{b} - T^{a}{}_{b c} \right) 
\label{contorDef}
\end{equation}%
is the {contortion tensor}.

Equation~(\ref{rela0alge}) is actually the content of a theorem, which states that any Lorentz connection can be decomposed into the spin connection of general relativity plus the contortion tensor \cite{HGV721}. The corresponding expression in terms of the spacetime--indexed linear connection reads
\be
\Gamma^\rho{}_{\mu\nu} = {\stackrel{\circ}{\Gamma}}{}^{\rho}{}_{\mu \nu} +
K^\rho{}_{\mu\nu},
\label{prela0}
\ee
where
\be
{\stackrel{\circ}{\Gamma}}{}^{\sigma}{}_{\mu \nu} = \onehalf \,
g^{\sigma \rho} \left( \partial_{\mu} g_{\rho \nu} +
\partial_{\nu} g_{\rho \mu} - \partial_{\rho} g_{\mu \nu} \right)
\label{lci}
\ee
is the zero--torsion Christoffel, or Levi--Civita connection, and 
\be
K^\rho{}_{\mu\nu} = {\textstyle
\frac{1}{2}} \left(T_\nu{}^\rho{}_\mu+T_\mu{}^\rho{}_\nu-
T^\rho{}_{\mu\nu}\right)
\label{contor}
\ee
is the spacetime--indexed contortion tensor.

\subsection{Behavior Under Lorentz Transformations}
\label{sec:LorentzTransf}
 
A local Lorentz transformation is fundamentally a transformation of the tangent space coordinates $x^a$:
\be
x'^a = \Lambda^a{}_b(x) \, x^b.
\ee
Under such a transformation, the tetrad transforms according to
\be
h'^{a} = \Lambda^{a}{}_b(x) \, h^b.
\ee
At each point of a riemannian spacetime, Eq.~(\ref{eq:tettomet}) only determines the tetrad up to transformations of the six--parameter Lorentz group in the tangent space indices. This means that there exists actually an infinity of tetrads $h_{a}{}^{\mu}$, each one relating the spacetime metric $g_{\mu \nu}$ to the tangent space metric $\eta_{c d}$ by Eqs.~(\ref{eq:gtoeta}) and (\ref{eq:tettomet}). In fact, any other Lorentz--rotated tetrad $\{h'_{a}\}$ will also relate the same metrics
\begin{equation}
g_{\mu \nu} = \eta_{c d}\, 
h'^{c}{}_{\mu} h'^{d}{}_{\nu}. 
\label{etatogmunu}
\end{equation} 
Under a local Lorentz transformation $\Lambda^{a}{}_{b}(x)$, the spin connection undergoes the transformation
\be
A'^{a}{}_{b \mu} = \Lambda^{a}{}_{c}(x) \, A^{c}{}_{d \mu} \,
\Lambda_{b}{}^{d}(x) +
\Lambda^{a}{}_{c}(x) \, \partial_{\mu} \Lambda_{b}{}^{c}(x).
\label{ltsc}
\ee
Of course, both curvature $R^a{}_{b \nu \mu}$ and torsion $T^a{}_{\nu \mu}$ transform covariantly:
\be
R'^{a}{}_{{b} \nu \mu} = \Lambda^{a}{}_{c}(x) \,\Lambda_b{}^d(x) \, R^c{}_{d \nu \mu}
\quad \mbox{and} \quad T'^{a}{}_{\nu \mu} = \Lambda^{a}{}_{b}(x) \, T^b{}_{\nu \mu}.
\ee

\subsection{Purely Inertial Lorentz Connection}
\label{InerGra}

In special relativity, Lorentz connections represent only inertial effects present in a given frame. In the class of inertial frames, where these effects are absent, the Lorentz connection vanishes identically. Since this is the class most used in field theory, Lorentz connections do not routinely show up in relativistic physics. Of course, as long as physics is frame independent, it can be described in any class of frames. For the sake of simplicity, however, one always uses the class of inertial frames when dealing with non--gravitational physics.

To see how an inertial Lorentz connection shows up, let us denote by $e^a{}_\mu$ a generic frame in Minkowski spacetime. The class of inertial (or holonomic) frames, defined by all frames  for which $f'^{c}{}_{a b} = 0$, will be denoted by $e'^a{}_\mu$. In a general coordinate system, the frames belonging to this class have the holonomic form
\be
e'^a{}_\mu = \partial_\mu x'^a,
\label{frame0}
\ee
with $x'^a$ a spacetime--dependent Lorentz vector: $x'^a = x'^a(x^\mu)$. The spacetime metric  
\be
\eta'_{\mu \nu} = e'^a{}_\mu e'^b{}_\nu \, \eta_{ab}
\label{eeee0}
\ee
still represents the Minkowski metric, but in a general coordinate system. In the specific case of cartesian coordinates, the inertial frame assumes the form
\be
e'^a{}_\mu = \delta_\mu^a
\label{frame00}
\ee
and the spacetime metric $\eta'_{\mu \nu}$ is that given by Eq.~(\ref{eq:etaofMinkoST}).
Under a local Lorentz transformation, the holonomic frame (\ref{frame0}) transforms according to
\be
e^a{}_\mu = \Lambda^a{}_b(x) \, e'^b{}_\mu.
\label{LoreTrans-e}
\ee
As a simple computation shows, it has the explicit form
\be
e^a{}_\mu = \partial_\mu x^a + \Aw^a{}_{b \mu} \, x^b \equiv \Dw_\mu x^a,
\label{InertiaTetrad}
\ee
where
\be
\Aw^a{}_{b \mu} = \Lambda^a{}_e(x) \, \partial_\mu \Lambda_b{}^e(x)
\label{InerConn}
\ee
is a Lorentz connection that represents the inertial effects present in the new frame $e^a{}_\mu$. As can be seen from Eq.~(\ref{ltsc}), it is just the connection obtained from a Lorentz transformation of the vanishing spin connection $\Aw'^e{}_{d \mu} = 0$:
\be
\Aw^a{}_{b \mu} = \Lambda^a{}_e(x) \, \Aw'^e{}_{d \mu} \, \Lambda_b{}^d(x) +
\Lambda^a{}_e(x) \, \partial_\mu \Lambda_b{}^e(x).
\ee 
Starting from an inertial frame, different classes of frames are obtained by performing {\em local} (point--dependent) Lorentz transformations $\Lambda^{a}{}_b(x^\mu)$. Inside each class, the infinitely many frames are related through {\em global} (point--independent) Lorentz transformations, $\Lambda^{a}{}_b =$~constant.

The inertial connection~(\ref{InerConn}) is sometimes referred to as the Ricci coefficient of rotation \cite{MTW732}. Due to its presence, the transformed frame $e^a{}_\mu$ is no longer holonomic. In fact, its coefficient of anholonomy is given by
\be
f^c{}_{a b} = - \left(\Aw^c{}_{a b} - \Aw^c{}_{b a} \right),
\ee
where we have used the identity $\Aw^a{}_{b c} = \Aw^a{}_{b \mu} \, e_c{}^\mu$. The inverse relation is
\be
\Aw^{a}{}_{b c} = \onehalf \left(f_{b}{}^{a}{}_{c} 
+ f_{c}{}^{a}{}_{b} - f^{a}{}_{b c} \right). 
\label{InerConefff}
\ee
Of course, as a purely inertial connection, $\,\Aw^a{}_{b \mu}$ has vanishing curvature and torsion:
\be
\Rw^{a}{}_{b \nu \mu} \equiv \partial_{\nu} \Aw^{a}{}_{b \mu} -
\partial_{\mu} \Aw^{a}{}_{b \nu} + \Aw^a{}_{e \nu} \, \Aw^e{}_{b \mu}
- \Aw^a{}_{e \mu} \, \Aw^e{}_{b \nu} = 0
\label{curvaDefW}
\ee
and
\be
\Tw^a{}_{\nu \mu} \equiv \partial_{\nu} e^{a}{}_{\mu} -
\partial_{\mu} e^{a}{}_{\nu} + \Aw^a{}_{e \nu} \, e^e{}_{\mu}
- \Aw^a{}_{e \mu} \, e^e{}_{\nu} = 0.
\label{tordefW}
\ee

\subsection{Equation of Motion of Free Particles}

As a concrete example, let us consider the equation of motion of a free particle. In the class of inertial frames $e'^a{}_\mu$, such particle is described by the equation of motion
\be
\frac{d u'^a}{d\sigma} = 0,
\ee
with $u'^a$ the particle four--velocity, and
\be
d \sigma^2 = \eta_{\mu \nu} \, dx^\mu dx^\nu
\label{MinkoInter}
\ee
the quadratic Minkowski invariant interval. In a anholonomic frame $e^a{}_\mu$, related to $e'^a{}_\mu$ by the local Lorentz transformation (\ref{LoreTrans-e}), the equation of motion assumes the manifestly covariant form
\be
\frac{d u^a}{d\sigma} + \Aw^a{}_{b \mu} \, u^b \, u^\mu = 0,
\label{anholoEM}
\ee
where
\be
u^a = \Lambda^a{}_b(x) \, u'^b
\ee
is the Lorentz transformed four--velocity, with
\be
u^\mu = u^a \, e_a{}^\mu
\ee
the spacetime--indexed four--velocity, which has the usual holonomic form
\be
u^\mu = \frac{d x^\mu}{d\sigma}.
\ee
Observe that the inertial forces coming from the frame non--inertiality are represented by the inertial connection of the left--hand side, which is of course non--covariant.

\section{General Relativity}
\label{InerEquations}

General Relativity conceives the gravitational interaction as a change in the geometry of spacetime itself. Specifically, as a change from the Lorentz metric $\eta_{\mu \nu}$ of Minkowski space into a riemannian metric $g_{\mu \nu}$. This new metric plays the role of basic field, and is in principe defined everywhere. Derivatives compatible with this overall presence of the same metric must preserve it, must parallel--transport it everywhere. Of all such Lorentz connections preserving $g_{\mu \nu}$, the most natural choice from the point of view of universality is to pick up the Christoffel, or Levi--Civita connection
\be
\Gammabol^\sigma{}_{\mu \nu} = \onehalf \,
g^{\sigma \rho} \left( \partial_{\mu} g_{\rho \nu} +
\partial_{\nu} g_{\rho \mu} - \partial_{\rho} g_{\mu \nu} \right),
\ee
which is a connection determined solely by the ten components of the metric tensor $g_{\mu \nu}$. It is the only metric--preserving connection with vanishing torsion, a magnitude which is then found not to play any role in the general--relativistic description of the gravitational interaction. The corresponding spin connection is
\be
\Abol^{a}{}_{b \mu} =
h^{a}{}_{\nu} \partial_{\mu}  h_{b}{}^{\nu} +
h^{a}{}_{\nu} \Gammabol^{\nu}{}_{\rho \mu} h_{b}{}^{\rho}.
\ee
Its Riemann curvature,
\be
\Rbol^{a}{}_{b \nu \mu} = \partial_{\nu} \Abol^{a}{}_{b \mu} -
\partial_{\mu} \Abol^{a}{}_{b \nu} + \Abol^a{}_{e \nu} \Abol^e{}_{b \mu}
- \Abol^a{}_{e \mu} \Abol^e{}_{b \nu},
\ee
represents the fundamental field of the theory: gravitation is present whenever at least one of its components is non--vanishing.

The field equation governing the dynamics of general relativity is Einstein equation
\be
\Rbol^a{}_{\nu} - {\textstyle{\frac{1}{2}}}\, \Rbol \, h^a{}_{\nu} = k \, \Theta^a{}_{\nu},
\label{EinsFiEq}
\ee
where $k = 8 \pi G/c^{4}$,
\begin{equation}
\Rbol^a{}_{\nu} = \Rbol^{\rho a}{}_{\rho \nu}
\quad \mbox{and} \quad \Rbol = h^a{}_{\nu} \Rbol_a{}^{\nu}
\label{RicciScalar}
\end{equation}
are, respectively, the Ricci and the scalar curvature, and
\begin{equation}
\Theta^a{}_{\nu} = -\, \frac{1}{\sqrt{-g}} \, \frac{\delta {\mathcal L}_s}{\delta h_a{}^{\nu}} 
\label{eq:Phys.8.[2.5]}
\end{equation}
is the symmetric source energy--momentum tensor modified by the presence of gravitation,with ${\mathcal L}_s$ the source field lagrangian. This equation can be obtained from the lagrangian
\be
{\mathcal L} = \Lbol + {\mathcal L}_s,
\label{eq:Phys.8.[1.12]}
\ee
where
\be
\Lbol = -\;  \frac{1}{2k} \, \sqrt{-g} \; \Rbol
\label{EinsHilberLag}
\ee
is the Einstein--Hilbert lagrangian of general relativity.

\subsection{How Does General Relativity Describe Gravitation?}

The curvature of the Levi--Civita connection gives rise to a geometric description of the gravitational interaction. To understand what such a {\em geometrical description}  does mean, let us consider the motion of a (spinless) particle in a gravitational field. In Minkowski spacetime, such a particle obeys the equation
\be
\frac{d u^a}{d\sigma} = 0,
\label{FreePartEqMot}
\ee
with $d\sigma$ the Minkowski invariant interval~(\ref{MinkoInter}). To obtain the equation valid in the presence of gravitation, a rule turns up, which is reminiscent of the gauge prescription: the minimal coupling prescription. According to this prescription, all ordinary derivatives must be replaced by covariant derivatives. In the specific case of general relativity, the free equation of motion (\ref{FreePartEqMot}) becomes the usual geodesic equation
\be
\frac{du^a}{ds} + 
\Abol^{a}{}_{b \nu} \, u^b u^{\nu} = 0,
\label{eq:geodesic}
\ee
where
\be
ds^2 = g_{\mu \nu} \, dx^\mu dx^\nu
\label{CurveInter}
\ee
is the riemannian spacetime quadratic interval. This equation describes the motion of a test particle in the presence of a gravitational field. It says essentially that the four--acceleration of the particle vanishes:
\be
{\stackrel{\circ}{a}}{}^{a} = 0.
\ee
This means that in General Relativity {\em there is no concept of gravitational force}. In this theory, the gravitational interaction is geometrized: the presence of gravitation produces a curvature in spacetime, and the gravitational interaction is described by letting particles to follow freely the spacetime curvature. 

\section{Teleparallel Gravity}
\label{sec:TeleGrav}

Teleparallel gravity corresponds to a gauge theory for the translation group. Accordingly, the gravitational field is represented by a translational gauge potential ${B^a{}_\mu}$, a 1-form assuming values in the Lie algebra of the translation group:
\be
B_\mu = B^a{}_\mu \, P_a.
\ee
It appears as the non--trivial part of the tetrad,
\be
h^a{}_\mu = e^a{}_\mu + B^a{}_\mu,
\label{TeleTetrada2}
\ee
with
\be
e^a{}_\mu \equiv \Dw_\mu x^a = \partial_\mu x^a + \Aw^a{}_{b \mu} \, x^b
\label{NonGraTetra}
\ee
the trivial (non--gravitational) tetrad. Under a gauge translation
\be
\delta x^a = \varepsilon^a,
\ee
the potential $B^a{}_\mu$ transforms according to
\be
\delta B^a{}_\mu = - \, \Dw_\mu \varepsilon^a.
\label{BamGauTrans}
\ee
The tetrad is consequently gauge invariant:
\be
\delta h^a{}_\mu = 0.
\ee

The field strength of teleparallel gravity is a 2-form assuming values in the Lie algebra of the translation group. In a general Lorentz frame its components are given by
\be
\Tw^a{}_{\mu \nu} = \partial_\mu B^a{}_\nu - \partial_\nu B^a{}_\mu +
\Aw^a{}_{b \mu} B^b{}_{\nu} - \Aw^a{}_{b \nu} B^b{}_{\mu},
\label{tfs}
\ee
or equivalently
\be
\Tw^{a}{}_{\mu \nu} = \Dw_\mu B^a{}_\nu - \Dw_\nu B^a{}_\mu.
\label{tfs1}
\ee
Since 
\be
\Dw_\mu \Dw_\nu x^a - \Dw_\nu \Dw_\mu x^a = 0,
\ee
it can be rewritten in the form
\be
\Tw^a{}_{\mu \nu} = \Dw_\mu h^a{}_\nu -
\Dw_\nu h^a{}_\mu.
\label{tfs3}
\ee
We see in this way that the field strength is nothing else, but torsion.
On account of the gauge invariance of the tetrad, the field strength is also invariant under gauge transformations:
\be
\Tw'^a{}_{\mu \nu} = \Tw^a{}_{\mu \nu}.
\ee
This is an expected result. In fact, considering that the generators of the adjoint representation are the coefficients of structure of the group taken as matrices, and considering that these coefficients vanish for abelian groups, fields belonging to the adjoint representations of abelian gauge theories will always be gauge invariant.

\subsection{Teleparallel Lorentz Connection}
\label{sec:BasicFields}

The fundamental Lorentz connection of teleparallel gravity is the purely inertial connection (\ref{InerConn}). This means that in this theory Lorentz connections keep the special--relativistic role of representing inertial effects only. Of course, as a purely inertial connection, its curvature vanishes identically:
\be
\Rw^{a}{}_{b \mu \nu} = \partial_{\mu} \Aw^{a}{}_{b \nu} -
\partial_{\nu} \Aw^{a}{}_{b \mu} + \Aw^a{}_{e \mu} \Aw^e{}_{b \nu}
- \Aw^a{}_{e \nu} \Aw^e{}_{b \mu} = 0.
\ee
However, for a tetrad involving a non--trivial translational gauge potential $B^a{}_\mu$, that is, for
\be
B^a{}_\mu \neq \Dw_\mu \varepsilon^a,
\label{NonTriB}
\ee
torsion will be non--vanishing:
\be
\Tw^a{}_{\mu \nu} = \partial_{\mu} h^{a}{}_{\nu} -
\partial_{\nu} h^{a}{}_{\mu} + \Aw^a{}_{e \mu} h^e{}_{\nu}
- \Aw^a{}_{e \nu} h^e{}_{\mu} \neq 0.
\ee
In teleparallel gravity, therefore, gravitation is represented by torsion, not by curvature. This is at variance with General Relativity, whose spin connection $\Abol^{a}{}_{b \mu}$ has vanishing torsion
\be
\Tbol^a{}_{\mu \nu} = \partial_{\mu} h^{a}{}_{\nu} -
\partial_{\nu} h^{a}{}_{\mu} + \Abol^a{}_{e \mu} h^e{}_{\nu}
- \Abol^a{}_{e \nu} h^e{}_{\mu} = 0,
\ee
but non--vanishing curvature
\be
\Rbol^{a}{}_{b \mu \nu} = \partial_{\mu} \Abol^{a}{}_{b \nu} -
\partial_{\nu} \Abol^{a}{}_{b \mu} + \Abol^a{}_{e \mu} \Abol^e{}_{b \nu}
- \Abol^a{}_{e \nu} \Abol^e{}_{b \mu} \neq 0.
\ee

The spacetime--indexed linear connection corresponding to the inertial spin connection (\ref{InerConn}) is
\be
\Gammaw^{\rho}{}_{\nu \mu} = h_{a}{}^{\rho} \partial_{\mu} h^{a}{}_{\nu} +
h_{a}{}^{\rho}\Aw^a{}_{b \mu} \, h^b{}_\nu \equiv
h_{a}{}^{\rho} \, \Dw_{\mu} h^{a}{}_{\nu}.
\label{gecow}
\ee
This is the so--called Weitzenb\"ock connection. Its definition is equivalent to the identity
\be
\partial_{\mu}h^a{}_{\nu} +
\Aw^a{}_{b \mu} \, h^b{}_\nu -
\Gammaw^{\rho}{}_{\nu \mu} \, h^a{}_{\rho} = 0.
\label{cacd1}
\ee
In the class of frames in which the spin connection $\Aw^a{}_{b \mu}$ vanishes, it reduces to
\be
\partial_{\mu}h^a{}_{\nu} -
\Gammaw^{\rho}{}_{\nu \mu} \, h^a{}_{\rho} = 0,
\label{cacd0}
\ee
which is the absolute, or distant parallelism condition, from where teleparallel gravity got its name. We notice finally that, for the specific case of the Weitzenb\"ock connection, identity~(\ref{prela0}) assumes the form
\be
\Gammaw^\rho{}_{\mu \nu} = \Gammabol^\rho{}_{\mu \nu} + \Kw^\rho{}_{\mu \nu},
\label{AwKbol}
\ee
where
\be
\Kw^\rho{}_{\mu \nu} = \onehalf \left(\Tw_\mu{}^\rho{}_\nu +
\Tw_\nu{}^\rho{}_\mu - \Tw^\rho{}_{\mu \nu} \right)
\label{STIContorDef}
\ee
is the contortion of the Weitzenb\"ock torsion
\be
\Tw^\rho{}_{\nu \mu} = \Gammaw^{\rho}{}_{\mu \nu} - \Gammaw^{\rho}{}_{\nu \mu}.
\label{TGaminusGa}
\ee

\subsection{Teleparallel Lagrangian}
 
The lagrangian density of teleparallel gravity is \cite{maluf94}
\begin{equation}
\Lw =
\frac{h}{4 k} \; \Tw_{\rho\mu\nu} \, \sw^{\rho\mu\nu},
\label{gala}
\end{equation}
where
\begin{equation}
\sw^{\rho\mu\nu} = - \, \sw^{\rho\nu\mu} =
\Kw^{\mu\nu\rho} - g^{\rho\nu}\,\Tw^{\sigma\mu}{}_{\sigma}
+ g^{\rho\mu}\,\Tw^{\sigma\nu}{}_{\sigma}
\label{S}
\end{equation}
is the so--called superpotential, with
\be
\Kw^{\nu}{}_{\rho \mu} = {\textstyle \frac{1}{2}} \left( \Tw_{\rho}{}^{\nu}{}_{\mu}
+ \Tw_{\mu}{}^{\nu}{}_{\rho} - \Tw^{\nu}{}_{\rho \mu} \right)
\ee
the contortion tensor of the teleparallel torsion. In terms of contortion it assumes the form
\begin{equation}
\Lw = {\frac{h}{2 k}} \left(\Kw^{\mu \nu \rho} \Kw_{\rho \nu \mu} -
\Kw^{\mu \rho}{}_\mu \Kw^\nu{}_{\rho \nu} \right).
\label{teleMoller}
\end{equation}
Substituting $\Kw^{\rho\mu\nu}$, we find
\be
\Lw =
{\frac{h}{2 k}} \left(\onefourth \; \Tw^\rho{}_{\mu\nu} \, \Tw_\rho{}^{\mu \nu} +
\onehalf  \; \Tw^\rho{}_{\mu\nu} \, \Tw^{\nu \mu}{}_\rho -
\, \Tw^\rho{}_{\mu\rho} \, \Tw^{\nu \mu}{}_\nu \right).
\label{TeleLagra}
\end{equation}
The first term corresponds to the usual lagrangian of internal gauge theories. The existence of the other two terms is related to the soldered character of the bundle. In fact, the presence of a tetrad field allows internal and external indices to be treated on the same footing, and consequently new contractions turn out to be possible.
In terms of algebraic--indexed torsion, the teleparallel lagrangian assumes the form
\be
\Lw =
{\frac{h}{2 k}} \left(\onefourth \; \Tw^a{}_{bc} \, \Tw_a{}^{bc} +
\onehalf  \; \Tw^a{}_{bc} \, \Tw^{c b}{}_a -
\, \Tw^a{}_{ba} \, \Tw^{c b}{}_c \right).
\label{TeleLagraBis}
\end{equation}
Notice that torsion is a Lorentz tensor~---~it transforms covariantly under local Lorentz transformations. It then follows that each term of this lagrangian is local Lorentz invariant, and consequently the whole lagrangian is also invariant independently of the numerical value of the coefficients.

\subsection{Equivalence with Einstein--Hilbert}
\label{sec:EquivWithEinsteinHilbert}
 
As we have already discussed, the curvature of the Weitzenb\"ock connection vanishes identically:
\be
\Rw^\rho{}_{\lambda\nu\mu} \equiv \partial_\nu \Gammaw^\rho{}_{\lambda \mu} -
\partial_\mu \Gammaw^\rho{}_{\lambda \nu} +
\Gammaw^\rho{}_{\eta \nu} \Gammaw^\eta{}_{\lambda \mu} -
\Gammaw^\rho{}_{\eta \mu} \Gammaw^\eta{}_{\lambda \nu} = 0. 
 \vspace{8pt}
\ee
Substituting the relation
\be
\Gammaw^{\rho}{}_{\mu \nu} =
{\stackrel{\circ}{\Gamma}}{}^{\rho} {}_{\mu \nu} +
\Kw^{\rho}{}_{\mu \nu},
\label{rela0bis}
\ee
we find
\be
\Rw^{\rho}{}_{\theta \mu \nu} \equiv
{\stackrel{\circ}{R}}{}^{\rho}{}_{\theta \mu \nu} +
\Qw^{\rho}{}_{\theta \mu \nu} = 0,
\label{relar}
\ee
where $\Rbol^{\rho}{}_{\theta \mu \nu}$ is the curvature of the Levi--Civita connection, and
\ba
\Qw^{\rho} {}_{\theta \mu \nu} =
\partial_{\mu}{}\Kw^{\rho}{}_{\theta \nu} - \partial_{\nu}{}\Kw^{\rho}{}_{\theta \mu}
+ \Gammaw^{\rho}{}_{\sigma \mu} \, \Kw^{\sigma}{}_{\theta \nu}
- \Gammaw^{\rho}{}_{\sigma \nu} \, \Kw^{\sigma}{}_{\theta \mu}~~~~~~~\quad \nonumber \\
- \, \Gammaw^{\sigma}{}_{\theta \mu} \, \Kw^{\rho}{}_{\sigma \nu}
+ \Gammaw^{\sigma}{}_{\theta \nu} \, \Kw^{\rho}{}_{\sigma \mu}
+ \Kw^{\rho}{}_{\sigma \nu} \, \Kw^{\sigma}{}_{\theta \mu}
- \Kw^{\rho}{}_{\sigma \mu} \, \Kw^{\sigma}{}_{\theta \nu} 
\label{qdk}
 \vspace{8pt}
\ea
is a tensor written in terms of the Weitzenb\"ock connection only. By taking appropriate contractions, the scalar version of identity (\ref{relar}) is found to be
\begin{equation}
-\,  \Rbol = \Qw \equiv \left(\Kw^{\mu \nu \rho} \Kw_{\rho \nu \mu} -
\Kw^{\mu \rho}{}_\mu \Kw^\nu{}_{\rho \nu} \right) +
\frac{2}{h} \, \partial_\mu \left(h \, \Tw^{\nu \mu}{}_\nu\right).
\end{equation}
Comparing with the teleparallel lagrangian~(\ref{teleMoller}), we see that
\begin{equation}
\Lw = \Lbol - \partial_\mu \Big(\frac{h}{k} \;
\Tw^{\nu \mu}{}_\nu \Big),
\label{LagraEquiva}
\end{equation}
where
\begin{equation}
\Lbol = -\, \frac{h}{2 k} \; \Rbol
\label{e-hl}
\end{equation}
is the Einstein--Hilbert lagrangian of general relativity. Up to a divergence, therefore,
the lagrangian of teleparallel gravity is equivalent to the lagrangian of general relativity.

To understand the presence of a divergence term between the two lagrangians, let us recall that the Einstein--Hilbert lagrangian (\ref{e-hl}) depends on the metric, as well as on the first and second derivatives of the metric. Equivalently, in the context of the tetrad formalism, we can say that it depends on the tetrad, as well as on the first and second derivatives of the tetrad field. The terms containing second derivatives, however, reduce to a divergence term \cite{landau}. In consequence, it is possible to rewrite the Einstein--Hilbert lagrangian in a form stating this aspect explicitly: 
\be
\Lbol = \Lbol_1 + \partial_\mu (\sqrt{-g} \, w^\mu),
\ee 
where $\Lbol_1$ is a lagrangian that depends solely on the tetrad and on its first derivatives, and $w^\mu$ is a four--vector. On the other hand, the teleparallel lagrangian (\ref{TeleLagraBis}) depends only on the tetrad and on its first derivative. The divergence in the equivalence relation (\ref{LagraEquiva}) is then necessary to remove the terms containing second derivatives of the tetrad from the Einstein--Hilbert lagrangian.

\subsection{Field Equations}
\label{sec:FieldEquations}

Consider now the lagrangian
\begin{equation}
{\mathcal L} = \Lw + {\mathcal L}_s,
\end{equation}
with ${\mathcal L}_s$ the lagrangian of a general source field. Variation with respect to the gauge potential $B^a{}_\rho$ ~---~or equivalently, with respect to the tetrad field $h^a{}_\mu$~---~yields the teleparallel version of the gravitational field equation \cite{livro2}
\be
\partial_\sigma(h \sw_a{}^{\rho \sigma}) -
k \, h \jw_{a}{}^{\rho} = k \, h\, {\Theta}_{a}{}^{\rho}.
\label{tfe10}
\ee
In this equation,  
\be
h \sw_a{}^{\rho \sigma} \equiv - \;
k \;  \frac{\partial {\Lw}}{\partial (\partial_\sigma h^a{}_{\rho})} =
\Kw^{\rho \sigma}{}_{a} - h_{a}{}^{\sigma} \,
\Tw^{\nu \rho}{}_{\nu} + h_{a}{}^{\rho} \, \Tw^{\nu \sigma}{}_{\nu}
\ee
is the superpotential, whereas the term 
\be
h \jw_{a}{}^{\rho} \equiv -\,
\frac{\partial \Lw}{\partial h^a{}_{\rho}} =
\frac{1}{k} \, h_a{}^{\mu} \, \sw_c{}^{\nu \rho} \,
\Tw^c{}_{\nu \mu} - \frac{h_a{}^{\rho}}{h} \, \Lw +
\frac{1}{k} \, \Aw^c{}_{a \sigma} \sw_c{}^{\rho \sigma}
\label{ptem10}
\ee
stands for the gauge current, which in this case represents the Noether energy--momentum density of gravitation itself \cite{gemt9}. Finally,
\begin{equation}
h\, {\Theta}_{a}{}^{\rho} = -\, \frac{\delta {\mathcal L}_s}{\delta h^a{}_{\rho}} \equiv -
\left( \frac{\partial {\mathcal L}_s}{\partial h^a{}_{\rho}} -
\partial_\mu \frac{\partial {\mathcal L}_s}{\partial_\mu \partial h^a{}_{\rho}} \right)
\label{memt1}
\end{equation}
is the source energy--momentum tensor.
Due to the anti--symmetry of the superpotential in the last two indices, the total --- that is, gravitational plus source --- energy--momentum density is conserved in the ordinary sense:
\be
\partial_\rho \big(h \jw_{a}{}^{\rho} + h\, {\Theta}_{a}{}^{\rho} \big) = 0.
\ee

The left--hand side of the gravitational field equation~(\ref{tfe10}) depends on the Weitzenb\"ock connection only. Using the identity~(\ref{rela0bis}), through a lengthy but straightforward calculation, it can be rewritten in terms of the Levi--Civita connection only:
\begin{equation}
\partial_\sigma \big(h\, \sw_a{}^{\rho \sigma}\big) -
k \, h \jw_{a}{}^{\rho} =
h\, \big({\stackrel{\circ}{R}}_a{}^{\rho} -
\onehalf \, h_a{}^{\rho} \,
{\stackrel{\circ}{R}} \big).
\label{ident}
\end{equation}
We see from this expression that, as expected due to the equivalence between the corresponding lagrangians, the teleparallel field equation (\ref{tfe10}) is equivalent to Einstein's field equation
\be
\Rbol_a{}^{\rho} -
\onehalf \, h_a{}^{\rho} \,
\Rbol = k \; {\Theta}_{a}{}^{\rho}.
\ee
Observe that the energy--momentum tensor appears as the source in both theories: as the source of curvature in general relativity, and as the source of torsion in teleparallel gravity. This shows that, according to teleparallel gravity, curvature and torsion are related to the same degrees of freedom of the gravitational field.

\subsection{How Does Teleparallel Gravity Describe Gravitation?}
\label{sec:PartGrav}

Let us take the geodesic equation of general relativity:
\be
\frac{du^a}{ds} +
\Abol^a{}_{b \nu} \, u^b \, u^\nu = 0.
\label{GeReGeoEqu}
\ee
Substituting the identity
\be
\Abol^a{}_{b \nu} = \Aw^a{}_{b \nu} - \Kw^a{}_{b \nu},
\ee
we obtain
\be
\frac{du^a}{ds} +
\Aw^a{}_{b \nu} \, u^b \, u^\nu = \Kw^a{}_{b \nu} \, u^b \, u^\nu.
\label{TeleForceEqu}
\ee
This is the teleparallel equation of motion of a particle of mass $m$ in a gravitational field --- as seen from a general Lorentz frame. It is a {\em force equation}, with contortion playing the role of gravitational force. The inertial forces coming from the frame non--inertiality are represented by the connection of the left-hand side, which is non-covariant by its very nature. In teleparallel gravity, therefore, whereas the gravitational effects are described by a covariant force, the non--inertial effects of the frame remain {\it geometrized} in the sense of general relativity, and are represented by an inertia--related Lorentz connection. Notice that in the geodesic equation~(\ref{GeReGeoEqu}), both inertial and gravitational effects are described by the connection term of the left--hand side.

Considering that the teleparallel force equation and the geodesic equation of general relativity are formally the same, the teleparallel description of the gravitational interaction is found to be equivalent to the description of general relativity. There are conceptual differences, though. In general relativity, a theory fundamentally based on the weak equivalence principle, curvature is used to {\it geometrize} the gravitational interaction. The gravitational interaction in this case is described by letting (spinless) particles to follow the curvature of spacetime. Geometry replaces the concept of force, and the trajectories are determined, not by force equations, but by geodesics. Teleparallel gravity, on the other hand, attributes gravitation to torsion. Torsion, however, accounts for gravitation not by geometrizing the interaction, but by acting as a force. In consequence, there are no geodesics in teleparallel gravity, only force equations quite analogous to the Lorentz force equation of electrodynamics \cite{AndPer976}. This is actually an expected result because, like electrodynamics, teleparallel gravity is a gauge theory.

\section{Final Remarks}

Although equivalent to general relativity, teleparallel gravity presents several distinctive features and achievements in relation to general relativity. For example, according to the geometric description of general relativity, which makes use of the torsionless Levi--Civita connection, there is a widespread belief that gravity produces a curvature {in spacetime}. In consequence, the Universe as a whole should be curved. However, the advent of teleparallel gravity breaks this paradigm. In fact, it becomes a matter of convention to describe the gravitational interaction in terms of curvature or in terms of torsion. This means that the attribution of curvature to spacetime is not an absolute, but a model--dependent statement. Here, we will discuss two additional points: the possibility of separating inertial effects from gravitation, and the existence of a true gravitational variable in the usual sense of classical field theory. 

\subsection{Separating Inertial Effects from Gravitation}
\label{SepaInerGrav}

Let us consider again the tetrad field
\be
h^a{}_\mu = \Dw_\mu x^a + B^a{}_\mu.
\label{Tetra6}
\ee
Whereas the first term on the right--hand side is purely inertial, the second is purely gravitational. This means that both inertia and gravitation are included in $h^a{}_\mu$. As a consequence, the coefficient of anholonomy of $h_a$,
\begin{equation}
f^c{}_{a b} = h_a{}^{\mu} h_b{}^{\nu} (\partial_\nu
h^c{}_{\mu} - \partial_\mu h^c{}_{\nu}),
\label{fcabBIS}
\end{equation}
will also represent both inertia and gravitation. Of course, the same is true for the spin connection of general relativity,
\begin{equation}%
\Abol^{a}{}_{b c} = \onehalf (f_b{}^a{}_c + f_c{}^a{}_b - f^{a}{}_{b c}). 
\label{tobetaken3}
\end{equation}%
In a local frame in which inertial effects exactly compensate gravitation, that connection vanishes,
\be
\Abol^{a}{}_{b c} \doteq 0,
\label{146}
\ee
and gravitation becomes locally undetectable. On the other hand, considering that the teleparallel spin connection represents inertial effects only, the identity
\be
\Abol^{a}{}_{b c} = \Aw^{a}{}_{b c} - \Kw^{a}{}_{b c}
\label{splittingPartimec}
\ee
corresponds actually to a separation of $\Abol^{a}{}_{b c}$ into inertial and gravitational parts \cite{Einstein056}. In fact, in the local frame in which (\ref{146}) holds, the identity (\ref{splittingPartimec}) becomes
\be
\Aw^{a}{}_{b c} \doteq \Kw^{a}{}_{b c}.
\ee
This expression shows explicitly that, in such a local frame, inertial effects (left--hand side) exactly compensate gravitation (right--hand side).

It is interesting to remark that, although the inertial part of $\Abol^{a}{}_{b c}$ does not contribute to some physical quantities, like curvature and torsion, it does contribute to others. An example is the energy--momentum density of gravitation, whose expression in general relativity always include, in addition to the energy--momentum density of gravity itself, also the energy--momentum density of inertial effects, which is non--tensorial by its very nature. This is the reason why in general relativity this density always shows up as a pseudotensor.\footnote{A sample of different pseudotensors can be found, for example, in Refs.~\cite{pseudo1, pseudo2, pseudo3, pseudo4, pseudo5, pseudo6, pseudo7, pseudo8, pseudo9}.} 
Furthermore, owing to its odd asymptotic behavior, the contribution of the inertial effects often yields unphysical (divergent or trivial) results for the total energy and momentum of a gravitational system. As a consequence, it is in general necessary to make use of a regularizing process to eliminate the spurious contribution coming from those inertial effects \cite{maluf10}. Due to the possibility of separating inertial effects from gravitation, in teleparallel gravity it is possible to write down a purely gravitational energy--momentum density which, as for any other field, is a true tensor. The existence of such a tensorial density allows one to compute unequivocally the energy and momentum of any gravitational system without necessity of a regularization process \cite{regW10}.

\subsection{A Genuine Gravitational Connection}
\label{GenuConne}

Due to the fact that spin connection of general relativity involves both gravitation and inertial effects, it is always possible to find a local frame in which inertial effects exactly compensate gravitation, in such a way that the connection vanishes at a point:
\be
\Abol^{a}{}_{b c} \doteq 0.
\ee
Since we know there is gravitation at that point, such connection is not a genuine gravitational variable in the usual sense of field theory. Notice, in particular, that any approach to quantum gravity based on this connection will necessarily include a quantization of the inertial forces --- whatever that may come to mean. Considering furthermore the divergent asymptotic behavior of the inertial effects, such approach will likely face additional difficulties.

Notice furthermore that the connection behavior of $\Abol^{a}{}_{b c}$ under local Lorentz transformations is due to its inertial content, not to gravitation itself. This can be seen from the decomposition (\ref{splittingPartimec}): whereas the first term on the right--hand side represents its inertial, non--covariant part, the second represents its gravitational part, which is a tensor. This means that it is not a genuine gravitational connection --- its gravitational content is covariant --- but just an inertial connection. One should not expect, therefore, any dynamical effect coming from a ``gaugefication'' of the Lorentz group. In this sense, local Lorentz transformations are similar to diffeomorphism, another symmetry empty of dynamical meaning. As a matter of fact, these two kind of transformations are used indistinctly in the metric formulation of general relativity, leading sometimes to the somewhat strange concept of ``locally inertial coordinate system''. This concept makes sense only if local Lorentz transformations between frames are considered on an equal footing with general coordinate transformations. Of course, this can be done as both transformations are empty of dynamical meaning.

In teleparallel gravity, on the other hand, the gravitational field is not represented by Lorentz connections, but by a translational--valued gauge potential $B^a{}_\mu$, the non--trivial part of the tetrad field. In this theory, Lorentz connections keep their special relativistic role, representing inertial effects only. Considering that the translational gauge potential represents gravitation only, to the exclusion of inertial effects, it cannot be made to vanish in a point through a choice of an appropriate frame. It is, for this reason, a true field variable in the usual sense of classical field theory. It is, furthermore, a genuine gravitational connection, and consequently the natural field--variable to be quantized in any approach to quantum gravity~\cite{livro2}.

\section*{Acknowledgments}
The author would like to thank R. Aldrovandi for useful discussions. He would like to thank also FAPESP, CAPES and CNPq for partial financial support.

\end{document}
